Microfluidic study of effects of flow velocity and nutrient concentration on biofilm accumulation and adhesive strength in a microchannel

N. LIU[1], T. Skauge[1], D. Landa-Marbán[3], B. Hovland[1], B. Thorbjørnsen[1], F. A. Radu[3], B. F. Vik[1], T. Baumann[4], G. Bødtker[1]

[1]Uni Research, Centre for Integrated Petroleum Research (CIPR), Realfagbygget, Allégaten 41, Bergen 5007, Norway.

[2]Department of Mathematics, Faculty of Mathematics and Natural Sciences, University of Bergen, Allégaten 41, 5020 Bergen, Norway.

[3]Institute of Hydrochemistry, Technische Universität München, Marchioninistr. 17, D-81377 München, Germany.

Corresponding author: Gunhild Bødtker (gubo@norceresearch.no)

Key Points:

- A T-shape microchannel was used to study biofilm morphology, accumulation and adhesive strength as responds to different velocities and nutrient concentrations by use of microscope.

- Optimized flow velocity ensures sufficient nutrients supplying with moderate shear stress for biofilm accumulation, while too high inhibits its formation.

- High nutrient concentration contributes to biofilm growth, but leads to a weak biofilm adhesive strength.

**Abstract**

Biofilm accumulation in the porous media can cause plugging and change many physical properties of porous media. Targeted bioplugging may have significant applications for industrial processes. A deeper understanding of the relative influences of hydrodynamic conditions including flow velocity and nutrient concentration, on biofilm growth and detachment is necessary to plan and analyze bioplugging experiments and field trials. The experimental results by means of microscopic imaging over a T-shape microchannel show that increase in fluid velocity could facilitate biofilm growth, but that above a velocity threshold, biofilm detachment and inhibition of biofilm formation due to high shear stress were observed. High nutrient concentration prompts the biofilm growth, but was accompanied by a relatively weak adhesive strength. This letter provides an overview of biofilm development in a hydrodynamic environment for better predicting and modelling the bioplugging associated with porous system in petroleum industry, hydrogeology, and water purification.

**Plain Language Summary**

In the recent decade, as the increasing requirement for green technologies, the use of bacteria has become more and more important in many applications. Bioplugging caused by bacteria growth in porous media might have some negative effects in industrial and medical applications because the clogging pores need extra cost to clean and prevention. However, engineering bioplugging has been explored as a viable technique for some applications, such as bioremediation, water purification and microbial enhanced oil recovery (MEOR). In order to control biofilms/biomasses selectively/directionally plugging in desirable places, the role of hydrodynamic conditions on biofilm growth and detachment is essential to investigate. Herein, a T-shape microchannel was prepared to study effects of flow velocity and nutrient concentration on biofilm accumulation and adhesive strength at pore scale. Our results suggest that flow velocity and nutrient concentration could control biofilm accumulation in both flowing and stagnant microchannels. The finding helps explain and predict the engineering bioplugging in porous media, especially for the selective plugging strategy of a MEOR field trial.

**1 Introduction**

Biofilm accumulation in porous media can cause bioplugging, leading to significant changes in physical properties of porous media, such as the reduction of porosity and permeability (Karambeigi et al., 2009; Karambeigi et al., 2013; Peszynska et al., 2016; Vilcaez et al., 2013). Up to now, applications of desired biofilm growth and its subsequent bioplugging have been attempted for various practices, such as in situ bioremediation (Joshi et al., 2017), soil injection (Oka & Pinder, 2017), waste treatment (Alhede et al., 2012; Manuel et al., 2007), ground water recharge (Brovelli et al., 2009) and microbial enhanced oil recovery (MEOR) (Karimi et al., 2012; Khajepour et al., 2014; Klueglein, et al. 2016; Rabiei et al., 2013; Sarafzadeh et al., 2014). In MEOR trails, biofilm accumulation leads to selective plugging of high permeability zones, subsequently forcing the diversion of injected fluids towards lower permeable zones to improve the oil recovery (Brown, 2010; Sarafzadeh et al., 2014). In order to understand and control selective bioplugging strategy, tremendous efforts have been taken in serval groups. Suthar et al. (2009) confirmed the obtained oil recovery because of bacterial growth and biofilm formation in the sand pack. (1, 2) Karambeigi et al. (2013) used two different heterogeneous micromodels to observe potential of bioplugging of high permeable layers of porous media for improving the efficiency of water flooding. Klueglein et al. (2016) investigated the influences of nutrients concentrations on cell growth and bioplugging potential during a MEOR trial. However, few works concern biofilm studies of its growth and detachment mechanisms accompanying the bioplugging process.

Bioplugging in porous media results from the accumulation of bacterial cells, production of extracellular polymeric substances (EPSs) in the pore space. Due to physicochemical properties of EPSS, biofilms can behave as viscous liquids to resist the flow-induced shear stress, and substantially plug the pore (Costerton et al., 1995; Flemming et al., 2011; Rozen et al., 2001; Stoodley et al., 1999; Tsai, 2005). Engineering bioplugging processes need to control biofilms selectively and substantially growing in desired places (Abdel Aal et al., 2010; Cuzman et al., 2015; Joshi et al., 2017). Therefore, mechanisms on biofilm development and its adhesive strength with solids surface is vitally important. It was found that biofilm growth and detachment could be significantly influenced by varying hydrodynamic conditions on the surrounding environment (Guimera et al., 2015; Rozen et al., 2001; Tsai, 2005). Biofilm growth and detachment rates could both increase with fluid velocity, as the increased mass transfer facilitating nutrients supply for bacteria growth, while the increased shear force in turn causing detachment (Lee et al., 2008; Stoodley et al., 1999; Tsai, 2005; Weiss et al., 2016). There is a consensus that biofilm growth rate increases with nutrients concentration, while nutrient starvation results in biofilm detachment (Cherifi et al., 2017; Hunt et al., 2004; Rochex & Lebeault, 2007). Nonetheless, knowledge on bioplugging must be depicted by examining a correlation between biofilm accumulations and its adhesive strength and hydrodynamic conditions like flow velocity and nutrient concentration, to improve understanding and hence ability to control bioplugging in fluid flooded porous systems.

Traditionally quiescent experiments for biofilm research were normally carried on homogeneous physical conditions, which lack environmental complexities for accurately determining the dynamic changes occurring during biofilm development (Rukavina & Vanic, 2016). The advent of new technologies, specially microfluidics, have attracted a rapidly growing interest to emulate biological phenomena by addressing unprecedented control over the flow conditions, providing identical and reproducible culture conditions, as well as real-time observation (Cherifi et al., 2017; Lee et al., 2008; Skolimowski et al., 2010; Tahirbegi et al., 2017). Indeed, microfluidics has been used for observing biofilm formation under various fundamental and applied researches, e.g. wastewater treatment (Raudales et al., 2014) and medical fields (Lam et al., 2016; Rozen et al., 2001). Herein, we used a T-shape microfluidic device equipped with a microscope to study the biofilm accumulation and adhesive strength as responds to various flow velocities and nutrient concentrations in the microchannel.

## 2 Materials and Methods

### 2.1 Bacteria and fluids

The bacteria used in the study was: Thalassospira strain A216101, a facultative anaerobic, nitrate-reducing bacteria (NRB), capable of growing under both aerobic and anaerobic conditions. It is able to grow on fatty acids and other organics acids as sole carbon and energy source. growth medium contained the following components($L^{-1}$): 0.02 g $Na_2SO_4$, 1.00 g $KH_2PO_4$, 0.10 g $NH_4Cl$, 20.00 g NaCl, 3.00 g $MgCl_2 \cdot 6H_2O$, 0.50 g KCl, 0.15 g $CaCl_2 \cdot 2H_2O$, 0.70 g $NaNO_3$, and 0.50 ml 0.20% resazurin. The medium is hereafter referred to as. After autoclaving in a dispenser, 1 L of growth medium was added 5 ml vitamin solution and 20 ml 1

M NaHCO$_3$ to adjust the pH to 6.80-7.20. Finally, pyruvate was added as the carbon source from a sterile stock solution to achieve final nutrient concentrations of 20 mM (2.0 N), 10 mM (1.0 N), 5 mM (0.5 N), and 1 mM (0.1 N), respectively. The final nutrient medium was stored at 4°C.

2.2 Experimental setup

The experimental apparatus is illustrated in Figure 1. A T-junction microfluidic device (Micronit, Netherland) consists of a single straight channel and a side channel with the sizes of 100 μm width and 20 μm depth and the nuzzle size at the cross-section as narrow as 10 μm (Figure S1). Bacterial inoculation was realized by injecting pre-cultured bacterial solution into the system from the bacterial injection channel (Channel 2). Then only nutrients with various pyruvate concentrations were injected from nutrients flow channel (Channel 1) at constant flowrates from 0.2 to 0.5 µl/min for approximately 6-7 days, while Channel 2 was closed, which led to a greater growing of bacteria on the substrates of the intersection of straight channel and side channel. The adhesion test was followed by inducing flow shear stress on biofilms through steadily increasing the flowrate to 1.2 µl/min. The corresponding flow velocity, Peclet number, Reynolds number and shear rate at each flowrate in Channel 1 are listed in Table S1.

2.3 Image process and effluent analysis

Image sequences on biofilm growth were acquired with a Leica microscope fitted with a digital camera (VisiCam 5.0, VWR) for scoring with time. The back illumination source is a cold halogen lamp with 24 V, 150 W. The main area of interest in this study is the intersection of straight channel and side channel, thereby two areas of interest (AOIs) with 0.5mm*0.1mm are extracted from the origin image for further image analysis (Figure S1 (b)). The image processing was performed using MATLAB®'s Image Processing Toolbox. Biofilm accumulation, here presented by biofilm coverage ($A_{nt}$) in areas of interest, was periodically measured in a flowing channel (Channel 1) and none-flowing Channel (Channel 2). Fluid samples were collected daily at the outlet through a quantitative real-time PCR (qPCR) on whole-cells to determine the total number of bacteria. Experiments were conducted at room temperature and pressure. Further details can be found in Support Information.

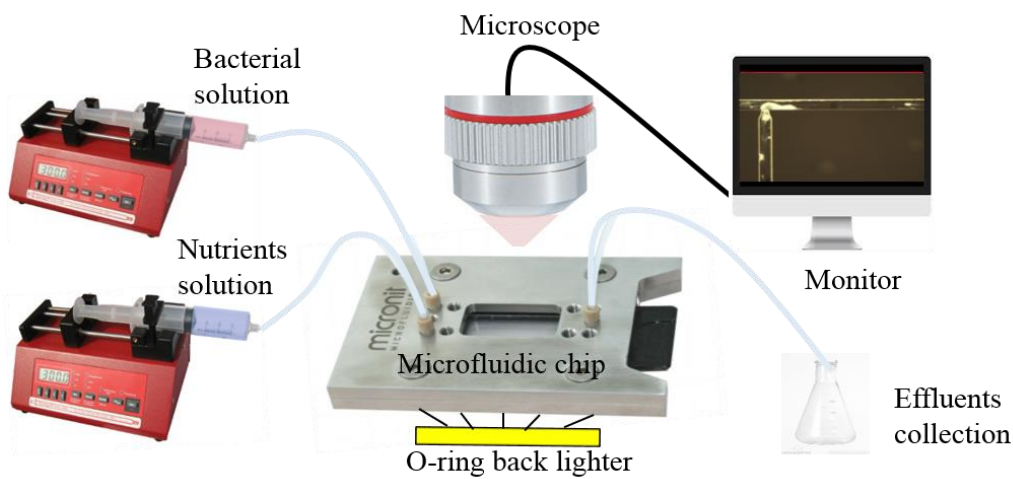

**Figure 1.** Schematic illustration of the experimental setup.

# 3 Results and discussion

## 3.1 Effect of flow velocity on biofilm accumulation and adhesive strength

Effects of flow velocity on biofilm development were measured by varying injecting flowrates of 10mM pyruvate (1.0 N, through Channel 1) from 0.2 µl/min to 0.5 µl/min, corresponding linear velocities from 1.66 to 4.17 mm/s respectively. After 6 days, the shear rate was steadily increased to 500.00 $s^{-1}$ to test the adhesive strength of biofilm attached on the solid surface. The accumulation of biofilms layer at different velocities was observed and registered as function of time by use of microscope.

**Table S1.** Table of Basic Flow Parameters at Various Flowrates in This Study

| Flowrate, µl/min | Water velocity, mm/s | Peclet number, Pe | Reynolds number, Re | Shear rate, $s^{-1}$ |
|---|---|---|---|---|
| 0.2 | 1.66 | 97.64 | 0.17 | 83.33 |
| 0.3 | 2.50 | 147.06 | 0.25 | 125.00 |
| 0.4 | 3.33 | 195.88 | 0.33 | 166.67 |
| 0.5 | 4.17 | 245.30 | 0.42 | 208.33 |
| 1.2 | 10.00 | 705.88 | 1.00 | 500.00 |

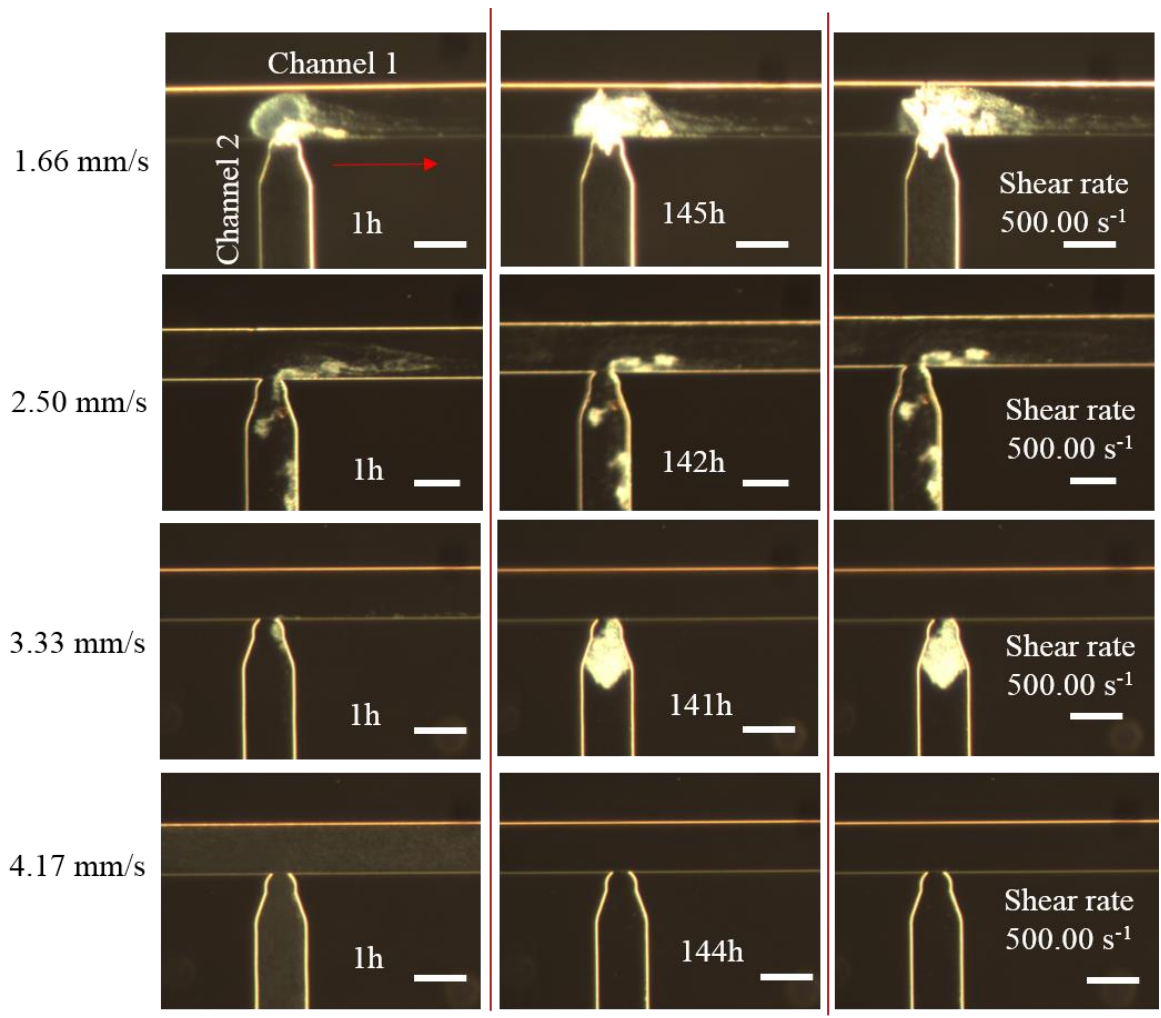

**Figure 2.** Optical images of biofilm growth in both microchannels at 1.0 N and various velocities. Images in the left column were taken after injecting nutrients for 1 h. The middle column shows images of biofilm growth for around 6 days. The right column lists images of biofilm detachment by increasing shear rate to 500.00 s$^{-1}$. Nutrients flow from right to left in the upper channel. Scale bars indicate 100 μm.

    Figure 2 shows images of biofilms development in two microchannels at various flow velocities. It is noticed that biofilms formed in Channel 1 reveal different morphological characters involving coverage and shape depending on the flow velocity. After inoculation, the initial attached biofilms at low velocities (1.66 and 2.50 mm/s) became irreversible and developed towards different structures along the nutrients flow. Biofilms at 1.66 mm/s tends to be approximately circular shape and has a larger coverage area, while biofilms at 2.50 mm/s show appearance of thin plate structures. On the contrary, there is no clear biofilm formation in Channel 1 at high velocities. These observations demonstrate that flow velocity has a direct effect on biofilm morphology.

    Biofilm growth in Channel 2 is highly dependent on the diffusion of nutrients in Channel 1. As the former bacteria injection path, most part of Channel 2 was full of biomasses without fluid shear forces. Only the void in the nozzle connecting with Channel 1 could act as the transport channel supplying nutrients for biofilm growth. As shown in Figure 2, biofilms at the

high shear rate of 166.67 s$^{-1}$ in Channel 1 led to a larger clusters compared with low rates, indicating that shear rate in Channel 1 determined the flux of nutrients transport to Channel 2. It is noticed that there was no biofilm growth in either channel at the highest flow velocity of 4.17 mm/s. This suggests that the high shear forces may prevent biofilm formation, which is in agreement with industrial applications where the formation of biofilm is prevented by high velocity flooding (Garrett et al., 2008).

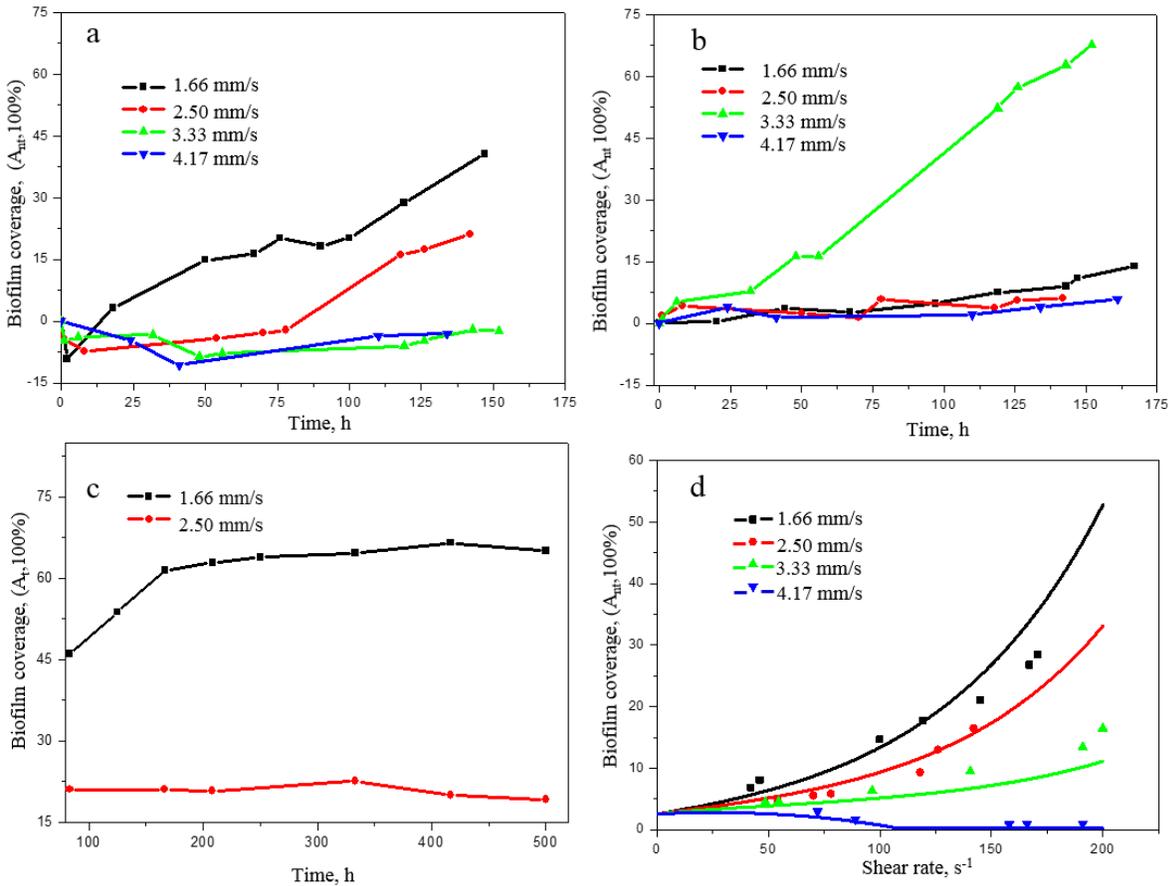

**Figure 3.** (a) Biofilm coverage over time in Channel 1 at various velocities; (b) Biofilm coverage over time in Channel 2 at different velocities; (c) Biofilm coverage in Channel 1 as response to the increasing shear rate after culturing biofilms at the velocities of 1.66 and 2.50 mm/s for 6 days; (d) Experimental data and numerical simulations of biofilm coverage in both channels at various velocities.

Figure 3 shows biofilm coverage as a function of time for different flow velocities in two microchannels. In this study, we set the initial biofilm coverage after inoculation to zero, and plot biofilm net coverage $A_{nt}$, by subtracting initial attachment to analysis biofilm accumulation. As shown in Figure 3 (a), the coverages of biofilm are under zero in Channel 1 in the early stage of injection, which demonstrates that the shear stress caused by nutrients flowing leads to snap-off of weak initial attachments. When the remained biofilms became irreversibly attached, they behaved as nuclei for new bacteria/biofilm growth, resulting in the increase of biofilm coverage.

As the velocity increased from 1.66 to 4.17 mm/s, biofilm coverage gradually decreased. This result illuminates that biofilm accumulation in the microchannel is highly related with flow velocities through two important factors, mass transfer and shear stress (Tsai, 2005; Weiss et al., 2016). As shown in Table S1, the Reynolds numbers in Channel 1 were very low (from 0.17 to 0.42), while the mass transfer Peclet number were extremely high (from 97.64 to 245.30), which suggests that mass transfer in the microchannel was dominated by convective actions and has negligible diffusion (Kirby, B., 2010). Thereby, the diffusion of nutrients from bulk to biofilms rarely increased when increasing the flow velocity, while the shear stress caused by water flow increased linearly. Thereby, the accumulation of biofilm, which is equal to its growth rate minus detachment rate, decreased with increasing flow velocities when shear stress induced detachment rate exceed growth rate.

Figure 3 (b) plots biofilm coverage in the bacterial injection channel (Channel 2) as a function of time in each run. It is noticed that biofilm accumulation in Channel 2 increased with shear rate in Channel 1 monotonically, which supports that the high shear rate increased the nutrient diffusive flux from Channel 1 to Channel 2, and facilitated biofilm growth in Channel 2. Therefore, for a confined no flowing system, biofilm accumulation rate is highly related to the nutrients availability, while the flow shear rate facilitates mass transfer, leading to an increase in biofilm accumulation. These observations are in correspondence with previous works (Cunningham et al., 1991; Rozen et al., 2001; Stoodley et al., 1999).

After 6 days of biofilm culturing, the shear rate steadily increased to 500.00 $s^{-1}$ to test the adhesive strength between biofilms and solid surfaces. As shown in the images in Figure 2, biofilms growth at 1.66 mm/s became elongated in the flowing direction to form filamentous "streamers" when the shear force acting on biofilms increasing with shear rate. Biofilm coverage as responds to increasing shear rate (Figure 3 (d)) shows that no large degree of detachment occurred in either experiments after 7 days. Compared to the large detachment at the initial stage, it suggests that the adhesive strength between biofilms and adhesive surface became stronger under shear (Billings et al., 2015; Flemming et al., 2011; Ohashi & Harada, 1994). Figure 3 (d) compares the experimental data with the mathematical model of biofilm coverages in both microchannels at various velocities, and shows that our experiment data is well fit with the numerical simulation.

3.2 Effect of nutrient concentration on biofilm accumulation and adhesive strength

To assess the influence of nutrient conditions on biofilm accumulation and adhesive strength, biofilms were grown at different nutrient concentrations. The baseline, 1.0 N, was 10 mM pyruvate in the growth medium and variations of two times (2.0 N), half (0.5 N) and one tenth (0.1 N) of the baseline concentration were applied. Injections were performed at a constant velocity of 1.66 mm/s from Channel 1 for approximately 7 days, and followed by a biofilm strength test by steadily increasing shear rate. The images are shown in Figure 4.

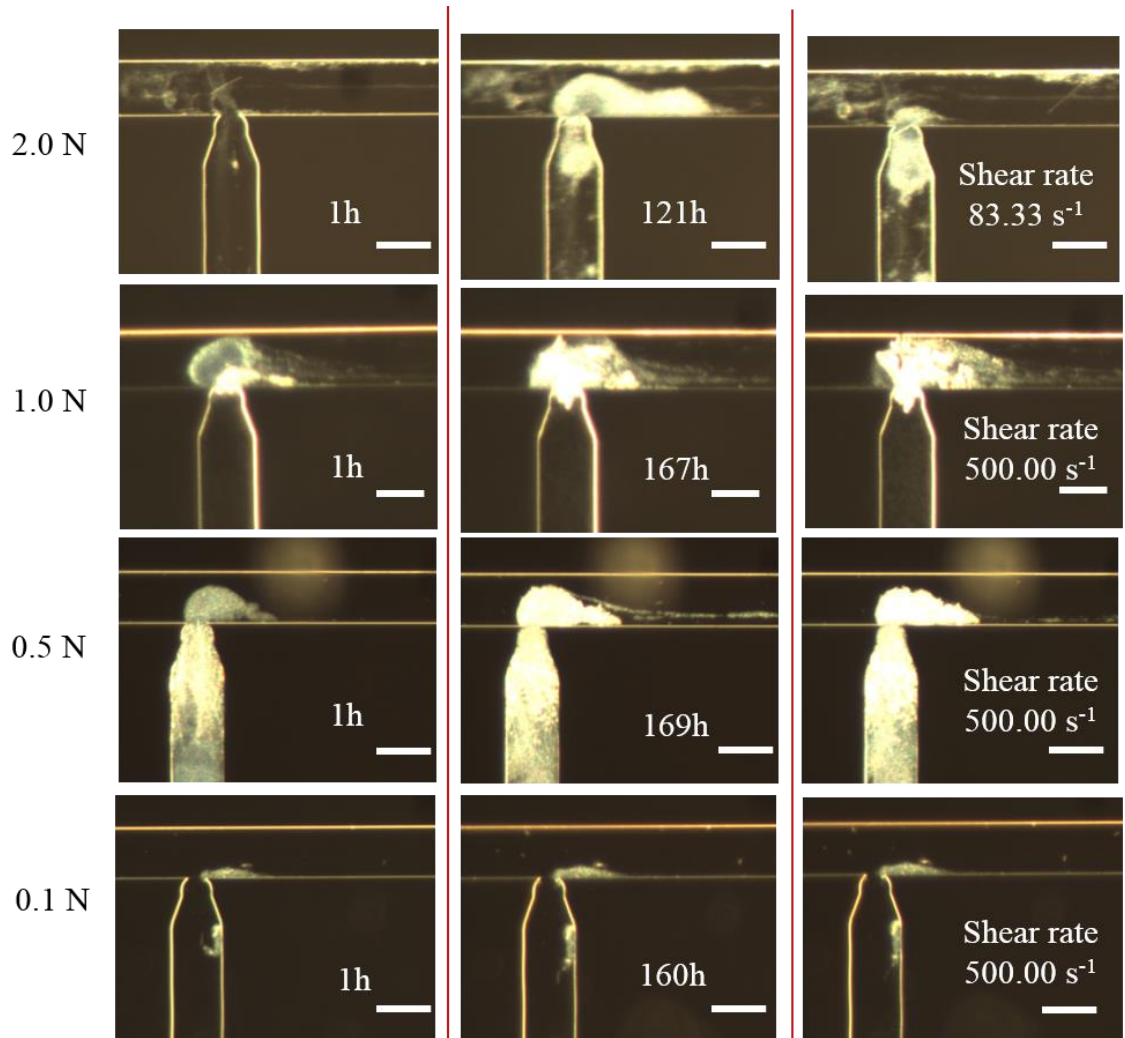

**Figure 4.** Optical images of biofilm growth over time at various nutrient concentrations, 2.0 N, 1.0 N, 0.5 N, and 0.1 N, respectively. Images in the left column were taken after injecting nutrients for 1 h. The middle column shows images of biofilm growth for around 7 days. The right column lists images of biofilm detachment by increasing shear rate to 500.00 s$^{-1}$. Nutrients flow from right to left in the upper channel. Scale bars indicate 100 μm.

    As shown in Figure 4, biofilm in Channel 1 with the highest concentration 2.0 N has a long, thick but loose structure, which is highly sensitive to the variation of shear stress. After 122 h, the formed biofilm was dispersed from the deep of the matrix, leaving behind a few attached biofilm spots to regrow. At nutrients input 1.0 N and 0.5 N, biofilm became denser and compacted, and the influence of shear stress reduced. The biofilm in Channel 2 at nutrient input 2.0 N also had a lower density than biofilm formed at lower nutrient concentrations, which further confirms that high nutrient concentrations lead to a low biofilm density. It is noticed that there is barely new biofilm formation at both channels at 0.1 N, which shows that the lowest nutrient input significantly limited biofilm growth and formation.

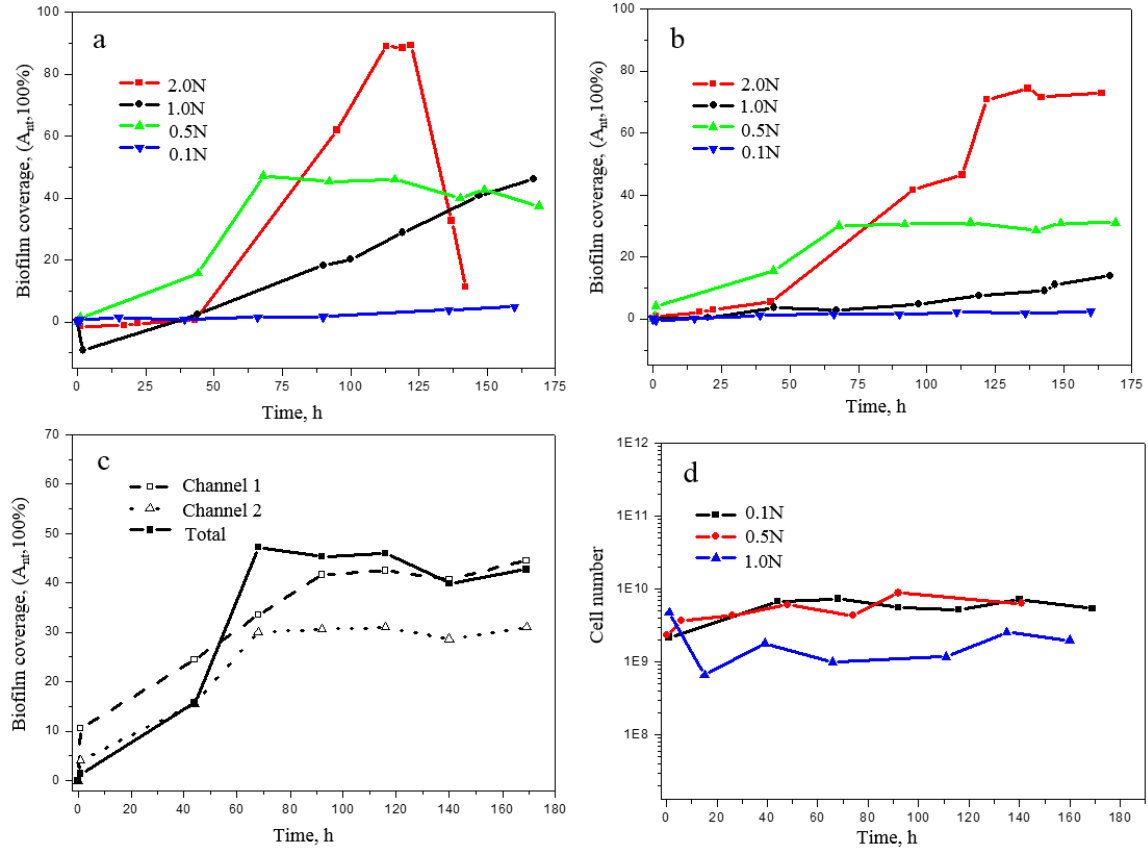

**Figure 5.** (a) Biofilm coverage over time in Channel 1 at different nutrient concentrations; (b) Biofilm coverage over time in Channel 2 at different nutrient concentrations; (c) Comparison of biofilm coverage in both channels at 0.5N; (d) Cell number of effluents at various nutrient concentrations.

Figure 5 shows biofilm coverage as a function of time for different nutrient concentrations in two microchannels. As shown in Figure 5 (a), in Channel 1, biofilm growth at a high nutrient concentration of 2.0 N has a much faster accumulation rate in the first 5 days, but rapidly decreased when it was detached from the substrate. Biofilm accumulation at the low nutrient concentration (0.5 N), is higher than that of 1.0 N in the first 3 days, but reached a plateau value after that, which was not observed for 1.0 N. This suggests that as biofilm grows in size, the increasing biomasses in the biofilm need more nutrients for growth, thereby the low nutrient concentration limits new biofilm formation. The lowest nutrient concentration (0.1 N) could not provide environment for biofilm growth. In this study, the limiting nutrient concentration for biofilm growth appears to be between 0.1 and 0.5 times N.

As shown in Figure 5 (b), biofilm accumulation in Channel 2 is influenced by nutrient concentrations. Biofilm formation at 2.0 N has larger coverage than other cases, indicating that high nutrients loading in Channel 1 leads to an increase in biofilm growth in Channel 2. In addition, stable plateau was obtained at later stages at 2.0 N than 0.5 N, suggesting that high nutrient concentration leads to a decrease in the time taken to reach the stable plateau in a no flow system. Figure 5 (c) shows that biofilm coverage obtained stable plateaus at 0.5 N in both channels. The time to reach the plateau in Channel 1 was later than that in Channel 2, indicating that flow shear rate can facilitate mass transfer and lead an increase in the time taken to reach the stable state. This result is further confirmed by cell number measurements from the effluent (shown in Figure 5 (d)). The cell numbers are relatively in the same level between 0.5 N and 1.0 N, which corresponds to high biofilm accumulation rates at 0.5 N and 1.0 N, suggesting that nutrient concentrations at this range are sufficient for bacterial growth and new biofilm formation. At 0.1 N, the cell number in the effluent decreased and no biofilm accumulated in the channel, indicating that less number of cells was released at limited nutrient loading.

It is noticed that biofilm growth at 2.0 N had a weak adhesive strength with substrates, because cells deep in the biofilm were dispersed from the interior of the biofilm matrix causing large degree of detachment. We observed this dispersion occurring at nutrient concentration of 2.0 N and flow velocities of 1.66 and 2.50 mm/s (Figure S2). Biofilms were observed to undergo growth and dispersion simultaneously at high nutrient concentrations (Figure S3). As biofilm growth at high rate at 2.0 N, cells trapped deeper in the biofilm matrix may have difficulties obtaining essential sources of energy or nutrients. In addition, waste products and toxins can accumulate fast in the biofilm community to reach toxic levels, threatening cells survival. Thus, microorganisms within the biofilm release from the matrix to resettle at a new location. In brief, biofilm growing under high nutrient concentration forms a loose structure with a high accumulation rate but a weak adhesive strength with substrates, which is easily detached by fluid shear. As nutrient concentration decreases, the biofilm accumulation rate decreases steadily and reaches a stable plateau when the nutrient loading is limited for biofilm growth (Flemming et al., 2011; Morgan et al., 2006; Wijman et al., 2007).

## 4 Conclusions

This work demonstrates that flow velocity and nutrient concentrations can have significant impact on biofilm accumulation in both flowing and stagnant microchannels. Negligible biofilm formation at the relatively high flow velocity of 4.17 mm/s and low nutrient concentration of 0.1 N suggests that there is a 'no/low growth region', where high shear forces lead to biofilm detachment and nutrient concentration is below the minimum required for biofilm formation. This is supported by the earlier work of Stoodley et al. (1999). At the conditions investigated in this work, a strong plugging effect in the microchannel were obtained at the relatively low flow velocity of 1.66 mm/s and the nutrient concentration of 1.0 N (10 mM substrate), which has a relative fast biofilms accumulation rate and a strong adhesion force to resist increase in flow-induced shear. This letter gives new insight to the relative influences of flowrate and nutrient concentration on biofilms development at pore scale. This may aid evaluations of bioplugging in porous systems such as for oil and ground water reservoirs. As potential permeability reducers in oil reservoirs, biofilm accumulation in porous media needs to be controlled by flow velocity and nutrient availability. Optimized nutrient flowrate ensures sufficient nutrients supplying rate with moderate shear stress in the microchannel, resulting in biofilm accumulation in both flowing and non-flow regions. However, too high stress may prevent biofilm formation and removal of adhered biofilms in the porous media. High nutrient concentration is beneficial for biofilm growth, but leads to a weak biofilm adhesive strength, which is easily detached by flow shear from the pores.

## Acknowledgments

We wish to thank Edin Alagic, Rikke H. Ulvøen and Tove L. Eide for technical assistance. This work was supported by the Research Council of Norway and industry partner GOE-IP through the projects IMMENS no. 255426.